\documentclass[letterpaper, 10 pt, journal, onside, twocolumn, nofonttune]{IEEEtran}
\IEEEoverridecommandlockouts
\usepackage[top=0.75in, left=0.75in, right=0.75in, bottom=0.75in, includefoot]{geometry}
\usepackage{mwe}
\usepackage{afterpage}

\usepackage{graphics} 
\usepackage{epsfig} 
\usepackage{color}
\usepackage[table,xcdraw,dvipsnames]{xcolor}

\usepackage{caption}
\usepackage{subcaption}

\usepackage{amsmath} 
\usepackage{nicefrac}
\usepackage{amssymb}  
\usepackage{bm}  
\usepackage{stmaryrd} 

\usepackage{amsthm}
\usepackage{bussproofs}
\theoremstyle{plain}

\theoremstyle{definition}

\theoremstyle{remark}

\usepackage{listings,comment}
\usepackage{algorithm}
\usepackage[noend]{algorithmic}

\usepackage{tikz}
\usepackage{pgfplots}
\pgfplotsset{compat=1.15}

\usepackage{forest} 

\usetikzlibrary{arrows,positioning,chains,fit,shapes,automata} 
\usepgfplotslibrary{patchplots}
\tikzset{
    >=stealth',
    punkt/.style={
           rectangle,
           rounded corners,
           draw=black, very thick,
           text width=12em,
           minimum height=2em,
           text centered},
    pil/.style={
           ->,
           thick,
           shorten <=2pt,
           shorten >=2pt,},
	cross/.style={
    	cross out, 
        draw=black, 
        minimum size=2*(#1-\pgflinewidth), 
        inner sep=0pt, 
        outer sep=0pt},
	cross/.default={1pt}
}

\makeatletter
\let\NAT@parse\undefined
\makeatother

\usepackage{hyperref}  
\usepackage[capitalise]{cleveref} 
\crefname{examplex}{Example}{Examples}
\crefname{definitionx}{Definition}{Definitions}
\crefname{problemx}{Problem}{Problems}

\usepackage[symbols,toc]{glossaries-extra}

\makenoidxglossaries

\setabbreviationstyle[acronym]{long-short}
\preto\chapter{\glsresetall}

\newacronym{cegis}{CEGIS}{Counterexample-Guided Inductive Synthesis}
\newacronym{csp}{CSP}{Constraint Satisfiability Problem}
\newacronym{cp}{CP}{Constraint Programming}
\newacronym{smt}{SMT}{Satisfiability Modulo Theories}
\newacronym{lp}{LP}{Linear Programming}
\newacronym{milp}{MILP}{Mixed-Integer Linear Programming}
\newacronym{ips}{IPS}{Intelligent Physical System}
\newacronym{ltl}{LTL}{Linear Temporal Logic}
\newacronym{rtl}{RTL}{Temporal Logic over Reals}
\newacronym{stl}{STL}{Signal Temporal Logic}
\newacronym{mpc}{MPC}{Model Predictive Control}
\newacronym{itmp}{ITMP}{Integrated Task and Motion Planning}
\newacronym{ai}{AI}{Artificial Intelligence}
\newacronym{ff}{FF}{fast forward}
\newacronym{idtmp}{IDTMP}{iteratively deepened task and motion planning}
\newacronym{cosmop}{CoSMoP}{composition of safe motion primitives}
\newacronym{mld}{MLD}{mixed logical dynamic}
\newacronym{pomdp}{POMDP}{partially observable Markov decision process}
\newacronym{prstl}{PrSTL}{probabilistic signal temporal logic}
\newacronym{socp}{SOCP}{Second-Order Cone Programming}
\newacronym{rhc}{RHC}{receding horizon control}
\newacronym{kf}{KF}{Kalman filter}
\newacronym{ukf}{UKF}{unscented Kalman filter}
\newacronym{ekf}{EKF}{extended Kalman filter}
\newacronym{smc}{SMC}{sequencial Monte-Carlo}
\newacronym{lqr}{LQR}{Linear Quadratic Regulator}
\newacronym{lqg}{LQG}{Linear Quadratic Gaussian}
\newacronym{zoh}{ZOH}{zero order hold}
\newacronym{ir}{IR}{infrared}
\newacronym{cps}{CPS}{cyber-physical system}
\newacronym{dof}{DOF}{degrees of freedom}
\newacronym{rrt}{RRT}{Rapidly-exploring Random Tree}
\newacronym{ltlopt}{LTLOpt}{optimal control with linear temporal logic specifications}
\newacronym{ros}{ROS}{Robot Operating System}
\newacronym{bsc}{BSC}{Bounded Satisfiability Checking}
\newacronym{ompl}{OMPL}{Open Motion Planning Library}
\newacronym{dfa}{DFA}{deterministic finite automata}
\newacronym{dba}{DBA}{deterministic Büchi automata}
\newacronym{iis}{IIS}{Irreducibly Inconsistent Set}
\newacronym{dnf}{DNF}{Disjunctive Normal Form}
\newacronym{bmc}{BMC}{Bounded Model Checking}
\newacronym{idrtl}{idRTL}{iterative deepening Real-time Temporal Logic}
\newacronym{sat}{SAT}{Satisfiability}
\newacronym{mlo}{MLO}{Maximum Likelihood Observation}

\makenoidxglossaries

\usepackage{xifthen}
\usepackage{xspace}

\makeatletter
\newcommand{\pushright}[1]{\ifmeasuring@#1\else\omit\hfill$\displaystyle#1$\fi\ignorespaces}
\newcommand{\pushleft}[1]{\ifmeasuring@#1\else\omit$\displaystyle#1$\hfill\fi\ignorespaces}
\makeatother

\DeclareMathOperator*{\argmax}{arg\,max}

\newcommand{\eye}[4]
{   \draw[rotate around={#4:(#2,#3)}] (#2,#3) -- ++(-.5*55:#1) (#2,#3) -- ++(.5*55:#1);
    \draw (#2,#3) ++(#4+55:.75*#1) arc (#4+55:#4-55:.75*#1);
    \draw[fill=gray] (#2,#3) ++(#4+55/3:.75*#1) arc (#4+180-55:#4+180+55:.28*#1);
    \draw[fill=black] (#2,#3) ++(#4+55/3:.75*#1) arc (#4+55/3:#4-55/3:.75*#1);
}

\newcommand{\mFormula}{\ensuremath{\varphi}}

\newcommand{\mAnd}{\ensuremath{\wedge}}
\newcommand{\mOr}{\ensuremath{\vee}}

\newcommand{\mUntil}{\ensuremath{\boldsymbol{U}}}
\newcommand{\mRelease}{\ensuremath{\boldsymbol{R}}}
\newcommand{\mAlways}{\ensuremath{\square}}
\newcommand{\mEventually}{\ensuremath{\Diamond}}

\newcommand{\mSat}{\ensuremath{\vDash}}

\definecolor{codegreen}{rgb}{0,0.6,0}
\definecolor{codegray}{rgb}{0.5,0.5,0.5}
\definecolor{codepurple}{rgb}{0.58,0,0.82}
\definecolor{backcolour}{rgb}{0.95,0.95,0.92}

\lstdefinestyle{mystyle}{
    backgroundcolor=\color{backcolour},   
    commentstyle=\color{codegreen},
    keywordstyle=\color{magenta},
    numberstyle=\tiny\color{codegray},
    stringstyle=\color{codepurple},
    basicstyle=\ttfamily\footnotesize,
    breakatwhitespace=false,         
    breaklines=true,                 
    captionpos=b,                    
    keepspaces=true,                 
    numbers=left,                    
    numbersep=5pt,                  
    showspaces=false,                
    showstringspaces=false,
    showtabs=false,                  
    tabsize=2
}

\def\BibTeX{{\rm B\kern-.05em{\sc i\kern-.025em b}\kern-.08em
    T\kern-.1667em\lower.7ex\hbox{E}\kern-.125emX}}
\begin{document}
\newgeometry{top=1in,left=0.75in,right=0.75in,bottom=0.75in, includefoot}
\afterpage{\aftergroup\restoregeometry}

\title{idSTLPy: A Python Toolbox for Active Perception and Control \\
\thanks{This work was supported in part by the National Science Foundation
under Grant IIS-1724070, Grant CNS-1830335 and Grant IIS-2007949.}
}

\author{Rafael Rodrigues da Silva$^1$ Kunal Yadav$^1$, and Hai Lin$^{1}$
	\thanks{$^{1}$ All authors are with Department of Electrical Engineering, University of Notre Dame, Notre Dame, IN 46556, USA.
		{\tt\small (rrodri17@nd.edu;~kyadav2@nd.edu;~hlin1@nd.edu)}}
}

\maketitle

\begin{abstract}
This paper describes a Python toolbox for active perception and control synthesis of probabilistic signal temporal logic (PrSTL) formulas of switched linear systems with additive Gaussian disturbances and measurement noises. We implement a counterexample-guided synthesis strategy that combines Bounded Model Checking, linear programming, and sampling-based motion planning techniques. We illustrate our approach and the toolbox throughout the paper with a motion planning example for a vehicle with noisy localization. The code is available at \url{https://codeocean.com/capsule/0013534/tree}.
\end{abstract}


\section{Introduction}

The recent decade has seen more intelligent systems in our day-to-day lives, but many of them are still pre-programmed or only work well in controlled environments.  Next-generation intelligent systems need to recognize their surrounding environments, make predictions of the environment behavior, and purposely take actions to improve confidence in their belief of environment states. This process is known as an active perception: the intelligent system explicitly explores the environment to collect more information about the environmental behavior \cite{5968}. \par

Since the process of active perception involves both actions and perceptions, we propose to specify an active perception task as \gls*{prstl} formulas, which combine real-time temporal logic with chance constraints. Then the active perception problem can be solved as a controller design for a given \gls*{prstl} specification with uncertain and differential constraints. 

Existing \gls*{prstl} controller synthesis methods include mixed-integer \gls*{socp} \cite{sadigh2015safe,zhong2017fast}, sampling-based optimization \cite{dey2016fast}, and heuristic-search based \cite{yoo2015control}. \gls*{socp} and sampling-based methods provide satisfying controllers for a convex fragment of \gls*{prstl}, but do not incorporate a perception model in the system dynamics. Thus these algorithms cannot synthesize controls to gather more information, and therefore are not considered active perception methods. 

In this paper, we introduce idSTLPy: a software toolbox for active perception and control developed based on our recent work in \cite{da2019active,dasilva2021active}. This toolbox is an open-source software package for designing the trajectory of an intelligent system with active perception from temporal specification and hybrid dynamics. Unlike other methods, this toolbox synthesizes controllers that consider the effects of observation on the belief dynamics. Hence, the planned trajectory includes motions that reduce the uncertainty about the state variables to achieve the task, i.e., active perception. 

Our current development is inspired by several toolboxes for symbolic control in the literature, such as TuLip \cite{7587949TuLip},
Linear Temporal Logic MissiOn Planning (LTLMoP) \cite{LTLMoP5650371} and Open Motion Planning Library \cite{he2015towards}, which support the design of controllers for deterministic hybrid systems from \gls*{ltl} formulas. However, to our best knowledge, idSTLPy would be the first toolbox that tackles active perception and control for stochastic systems.

The current version of idSTLPy models the stochastic system behavior as a switched linear system with Gaussian noises. This model allows us to inherit the computational efficiency and soundness of Kalman filtering. Additionally, these systems help to represent complex behaviors of physical systems interacting with logical rules or controllers. Therefore, these switched systems allow us to model several real-life problems. \par

The software is written in Python. Our basic idea is to combine \gls*{bmc} with sampling-based motion planning to separate logical and dynamical constraints. We propose abstractions that approximate the belief dynamics during the planning and permit us to use these techniques. We show through a simple example that the system can track the planned trajectory during the execution. Therefore, the main goal of this paper is to introduce the newly developed toolbox through a motion planning example under uncertain localization.

This paper is structured as follows: \cref{sec:preliminaries} briefly describes the preliminaries, \cref{sec:overview,sec:user-guide} gives the overview of the toolbox with an example. Finally, \cref{sec:conclusion} concludes the work.

\section{Preliminaries}\label{sec:preliminaries}

\subsection{System}\label{sec:prsystem}

We consider switched linear control systems as follows:
\begin{equation}\label{eq:prsystem}
    \begin{aligned}
	    \boldsymbol{x}_{k+1} = & A_{q_k} \boldsymbol{x}_k + B_{q_k} \boldsymbol{u}_k + W_{q_k} \boldsymbol{W}_k, & \boldsymbol{W}_k \sim \mathcal{N}(0, I)\\
	    \boldsymbol{y}_k = & C_{q_k} \boldsymbol{x}_k + n_{q_k}(\boldsymbol{x}_k) \boldsymbol{V}_k, & \boldsymbol{V}_k \sim \mathcal{N}(0, I),
    \end{aligned}
\end{equation}
where $\boldsymbol{x}_k \in \mathbb{R}^n$ are the state variables, $\boldsymbol{u}_k \in \mathcal{U} \subseteq \mathbb{R}^m$ are the input variables, $\mathcal{U} \subseteq \mathbb{R}^m$ is a polytope, $\boldsymbol{y}_k \in \mathbb{R}^p$ are the output variables. Each system location $q \in Q = \{1, 2, \dots, N \}$ is defined by a noise function $n_q : \mathbb{R}^n \rightarrow \mathbb{R}^n$, and constant matrices $A_q \in \mathbb{R}^{n \times n}$, $B_q \in \mathbb{R}^{n \times m}$, and $C_q \in \mathbb{R}^{p \times n}$ with proper dimensions. We assume that the system is subject to mutually uncorrelated zero-mean stationary Gaussian additive disturbances $\boldsymbol{V}_k \sim \mathcal{N}(0, I_n)$ and $\boldsymbol{W}_k \sim \mathcal{N}(0, I_p)$, where $I_n$ is the identity matrix with dimension $n$. Note that this dynamical system can arise from linearization and sampling of a more general continuous system. In such a case, we denote the sampling period as $T_s$, where $T_s = t_{k+1} - t_k $ for all $k \in \mathbb{N}_{\geq 0}$. We assume that the uncertainty is stable, meaning that the uncertainty does not increase infinitely over time.

\subsection{Trajectory}

The system in \cref{eq:prsystem} is probabilistic. This means that the dynamics result into a random process $\boldsymbol{X}_k$ that represent probabilities over the state variables $prob(\boldsymbol{X}_k = \boldsymbol{x}_k$). We call this random process as belief state. A belief trajectory $\boldsymbol{\beta}$ is defined as a sequence $\boldsymbol{X}_0 \xrightarrow{q_0, \boldsymbol{u}_0, \boldsymbol{y}_1} \boldsymbol{X}_1 \dots$. A transition $\boldsymbol{X}_k\xrightarrow{q_k, \boldsymbol{u}_k, \boldsymbol{y}_k}\boldsymbol{X}_{k+1}$ represents the process of applying a command $q_k \in Q$ and input $\boldsymbol{u}_k \in \mathcal{U}$ at instant $k$ and waiting for an observation $\boldsymbol{y}_{k+1}$ at instant $k + 1$ to update the next belief state $\boldsymbol{X}_{k+1}$. 

\subsection{Probabilistic Signal Temporal Logic}

We specify the requirements of a system belief trajectory using 
\gls*{prstl} formulas. These formulas are defined recursively according to the following grammar:
\begin{align*}
    \phi := & \pi^\mu_\epsilon | \pi^\mathbb{Q}  | \pi^{\mathbb{Q}_1} \mOr \pi^{\mathbb{Q}_2} | \phi_1 \mAnd \phi_2 \\
    \mFormula := & \phi | \mFormula_1 \mAnd \mFormula_2 | \mFormula_1 \mOr \mFormula_2 |  \mFormula_1 \mUntil_{[a,b]} \mFormula_2 |  \mFormula_1 \mRelease_{[a,b]} \mFormula_2,
\end{align*}
where $\pi$ is a predicate, $\mFormula$, $\mFormula_1$, and $\mFormula_2$ are \gls*{prstl} formulas, and $\phi$, $\phi_1$, and $\phi_2$ are  \gls*{prstl} state formulas. Predicates can be one of two types: atomic and probabilistic. An atomic predicate $\pi^\mathbb{Q}$ is a statement about the system locations and is defined by a subset $\mathbb{Q} \subseteq Q$ of locations. A probabilistic predicate $\pi^\mu_\epsilon$ is a statement about the belief $\boldsymbol{X}_k$ defined by a linear function $\mu : \mathbb{R}^n \rightarrow \mathbb{R}$ and a a tolerance $\epsilon \in [0,0.5]$. The operators $\mAnd, \mOr$ are Boolean operators conjuntion, and disjunction, respectively. The temporal operators $\mUntil$ and $\mRelease$ are \gls*{ltl} operators until and release, respectively. In \gls*{prstl}, these operators are defined by an interval $[a,b] \subseteq \mathbb{N}_{\geq 0}$. We assume that \gls*{prstl} state formulas forms a full-dimensional region in the state space $\mathbb{R}^n$. 

We denote the fact that a belief trajectory $\boldsymbol{\beta}$ satisfies an \gls*{prstl} formula $\mFormula$ with $\boldsymbol{\beta} \mSat \mFormula$. Furthermore, we write $\boldsymbol{\beta} \mSat_k \mFormula$ if the trajectory $\boldsymbol{X}_k \xrightarrow{q_k, \boldsymbol{u}_k, \boldsymbol{y}_{k+1}} \boldsymbol{X}_{k + 1} \dots$ satisfies $\mFormula$.  Formally, the following semantics define the validity of a formula $\mFormula$ with respect to the trajectory $\boldsymbol{\beta}$:
\begin{itemize}
  \item $\boldsymbol{\beta} \mSat_k \pi^\mathbb{Q}$ if and only if $k=0$ or $q_{k-1} \in \mathbb{Q}$,
  \item $\boldsymbol{\beta} \mSat_k \pi^\mu_\epsilon$ if and only if $p\big(\mu(\boldsymbol{x}_k) \leq 0\big) \geq 1 - \epsilon$,
  \item $\boldsymbol{\beta} \mSat_k \mFormula_1 \mAnd \mFormula_2$ if and only if $\boldsymbol{\beta} \mSat_k \mFormula_1$ and $\boldsymbol{\beta} \mSat_k \mFormula_2$,
  \item $\boldsymbol{\beta} \mSat_k \mFormula_1 \mOr \mFormula_2$ if and only if $\boldsymbol{\beta} \mSat_k \mFormula_1$ or $\boldsymbol{\beta} \mSat_k \mFormula_2$,
    \item $\boldsymbol{\beta} \mSat_{k} \mFormula_1 \mUntil_{[a,b]} \mFormula_2$ if and only if $\exists k^\prime$ s.t. $k+a \leq k^\prime \leq k+b$, $\boldsymbol{\beta} \mSat_{{k^\prime}}\mFormula_2$, and $\boldsymbol{\beta} \mSat_{{k^{\prime\prime}}}\mFormula_1$, $\forall k + a \leq k^{\prime\prime} < k^\prime$;
    \item $\boldsymbol{\beta} \mSat_{k} \mFormula_1 \mRelease_{[a,b]} \mFormula_2$ if and only if $\exists k^\prime$ s.t. $k+a \leq k^\prime \leq k+b$, $\boldsymbol{\beta} \mSat_{{k^\prime}}\mFormula_1$, and $\boldsymbol{\beta} \mSat_{{k^{\prime\prime}}}\mFormula_2$, $\forall k + a \leq k^{\prime\prime} \leq k^\prime$, or $\boldsymbol{\beta} \mSat_{{k^\prime}}\mFormula_2$, $\forall  k + a \leq k^\prime \leq k+b$,
  \item $\boldsymbol{\beta} \mSat \mFormula$ if and only if $\boldsymbol{\beta} \mSat_0 \mFormula$,
\end{itemize} 
where the temporal operators are indexed by its delay $a \in \mathbb{N}_{\geq 0}$ and deadline $b \in \mathbb{N}_{\geq 0}: a < b \leq \infty$. We can derive other operators such as \emph{true} ($\top = \pi^Q$), \emph{false} ($\perp = \pi^\emptyset$), \emph{always} ($\mAlways_{[a,b]} \mFormula = \perp \mRelease_{[a, b]} \mFormula$) and eventually ($\mEventually_{[a,b]} \mFormula = \top \mUntil_{[a, b]} \mFormula$).

\subsection{Problem Formulation}

A practical problem definition for active perception and control synthesis from \gls*{prstl} specification is a feasibility problem of the form,
\begin{equation}
\begin{aligned}
    \text{find } & \boldsymbol{\xi} \\
    \text{s.t. } & \boldsymbol{\xi} \mSat \mFormula, \\
    & prob(\boldsymbol{X}_0 = \bar{\boldsymbol{x}}) \sim \mathcal{N}(\bar{\boldsymbol{x}}, \bar{\Sigma}^x), \\
    & \boldsymbol{X}_{k+1} = A_{q_k} \boldsymbol{X}_k + B_{q_k} \boldsymbol{u}_k + W_{q_k}\boldsymbol{W}_k, \\
    & \boldsymbol{Y}_{k} = C_{q_k} \boldsymbol{X}_k + n_{q_k}(\boldsymbol{x}_k)\boldsymbol{V}_k, \\
    & \boldsymbol{y}_{k+1} = \argmax_{\boldsymbol{y}_{k+1}}prob(\boldsymbol{Y}_{k+1} = \boldsymbol{y}_{k+1}| \boldsymbol{X}_k, q_k, \boldsymbol{u}_k), \\
    & q_k \in Q, \boldsymbol{u}_k \in \mathcal{U}, \boldsymbol{W}_k \sim \mathcal{N}(0, I_n), \boldsymbol{V}_k \sim \mathcal{N}(0, I_p),
\end{aligned}
\end{equation}
where $\boldsymbol{\xi}$ is a belief trajectory, $\mFormula$ is a \gls*{prstl} formula, $prob(\boldsymbol{X}_0 = \bar{\boldsymbol{x}})$ is the initial condition (a priori belief), and $\argmax_{\boldsymbol{y}_{k+1}}prob(\boldsymbol{Y}_{k+1} = \boldsymbol{y}_{k+1}| \boldsymbol{X}_k, q_k, \boldsymbol{u}_k)$ is a practical approximation called \gls*{mlo} \cite{platt2010belief,dasilva2021active}.

\section{idSTLPy Overview}\label{sec:overview}

\begin{figure}
    \tikzstyle{block} = [draw, rounded corners=1mm, color=blue, text=black, line width=0.5mm, rectangle, minimum height=3em, minimum width=6em]
    \centering
    \begin{tikzpicture}[auto, >=latex', scale=0.75, transform shape]
        \node[block, minimum width=11em] (det) {Construct an Approximated System};
        \draw[->] ([yshift=1cm] det.north) -- node [pos=0.5, right] {$\text{System } (\ref{eq:prsystem}), \mathcal{N}(\bar{\boldsymbol{x}}, \bar{\Sigma}^x), \mFormula$} (det.north);
        \node[block, minimum width=11em,below=1.25cm of det] (abs) {Abstract the Approximated System and the Specification};
        \draw[->] (det.south) -- node [pos=0.5, right] {$\widehat{TS}, \mFormula$} (abs.north);
        \node[block, minimum width=11em,below=1.25cm of abs] (dplan) {\begin{tabular}{c}Bounded Model Checking (BMC) \\ $\widetilde{TS} \times \breve{TS}_{fair,1} \times \dots \times \breve{TS}_{fair,N} \mSat \boldsymbol{E} (\mFormula)_{LTL}$
        \end{tabular}};
        \draw[->] (abs.south) -- node [pos=0.5, right] {$\widetilde{TS}, (\mFormula)_{LTL}$} (dplan.north);
        \node[block, minimum width=11em,below=1.25cm of dplan] (reachsearch) {Dynamical Feasibility Check};
        \draw[->] (det.south) |- ++(-4.5cm,-0.5cm) |- (reachsearch);
        \draw[->,color=blue] ([xshift=-14mm] dplan.south) -- node[left,near end] {\color{blue} Example $\tilde{\boldsymbol{\beta}}_{K,L}$} node[right,pos=0.1] {\color{blue} \textbf{sat}} ([xshift=-14mm] reachsearch.north);
        \draw[->,color=red] ([xshift=10mm]reachsearch.north) -- node[right,pos=0.85] {\color{red} Counter-example $\widetilde{TS}_{cex,i}$} node[left,near start] {\color{red} \textbf{infeas}} ([xshift=10mm] dplan.south);
        \draw[->,thick,color=blue] (reachsearch.east) -- node[right, pos=1] {\color{blue} \begin{tabular}{l}
             \textbf{a trajectory that} \\
             \textbf{satisfies the specification} 
        \end{tabular}} ++(0.5cm,0);
        \draw[->,thick,color=red] (dplan.east) -- node[right,pos=1] {\color{red} \textbf{No solution}} ++(0.5cm,0);
    \end{tikzpicture}
    \caption{Pictorial representation of proposed approach. }
    \label{fig:diag1}
    \vspace{-0.5cm}
 \end{figure}
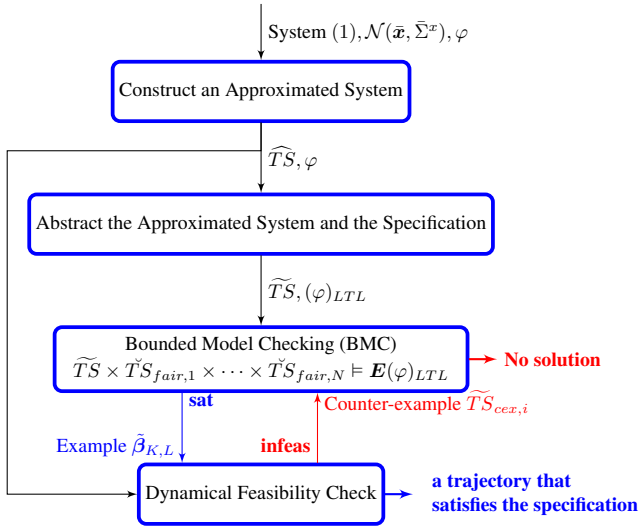

Our toolbox implements the approach in \cite{dasilva2021active} illustrated in \cref{fig:diag1}. The basic idea is to construct deterministic abstractions (i.e., $\widehat{TS}$ and $\widetilde{TS}$) and to use \textit{counterexample-guided synthesis} \cite{alur2013syntax,reynolds2015counterexample} to satisfy both the \gls*{prstl} specification $\mFormula$ and the dynamics of System (\ref{eq:prsystem}). Two interacting layers, discrete and continuous, work together to overcome nonconvexities in the logical specification $\mFormula$ efficiently. At the discrete layer, a discrete planner acts as a \textit{proposer}, generating discrete plans by solving a \gls*{bmc} \cite{biere2006linear,dasilva2021automatic}  for the given specification (i.e., $(\mFormula)_{LTL}$): $\widetilde{TS} \times \breve{TS}_{fair,1} \times \dots \times \breve{TS}_{fair,N} \mSat \boldsymbol{E} (\mFormula)_{LTL}$. We use an iterative deepening search to search first for shorter satisfying plans, thus minimizing undue computation. We pass the satisfying discrete plans to the continuous layer, which acts as a \textit{teacher}. In the continuous layer, a sampling-based search is applied to check whether a discrete plan is feasible. If the feasibility test does not pass, we construct a counterexample (i.e., $\breve{TS}_{fair,i}$) to discard infeasible trajectories. Then we add this counterexample to the discrete planner and repeat this process until we find a solution or no more satisfying plans exist. 

In this approach, we proposed a SPARSE-RRT \cite{littlefield2013efficient} --a sampling-based motion planning-- variant for active perception. The execution of this method is defined by a timeout in seconds ($rrt\_timeout$), a distance to consider that two states are near ($delta\_near$), a distance to drain near states ($delta\_drain$), a goal bias ($goal\_bias$), a minimum ($min\_num\_of\_steps$) and a maximum ($max\_num\_of\_steps$) number of steps for each iteration. Intuitively, for each candidate solution, we execute the proposed RRT for $rrt\_timeout$ seconds. During the execution, we randomly sample a state and take an existing trajectory that the last state is sufficient near ($delta\_near$) and has less uncertainty (i.e., active perception). Next, we randomly select a target state with probability $goal\_bias$ to be in the goal (i.e., task planning) and synthesize control inputs for an horizon between $min\_num\_of\_steps$ and $max\_num\_of\_steps$.  If the new trajectory has the last state with less uncertainty than other near ($delta\_drain$) trajectories, we remove the latter from the existing trajectory. If we find at least one trajectory that satisfies the \gls*{prstl} formula, we finish the search. Otherwise, we create a counter-example and find another candidate solution.  

\section{User Guide for idSTLPy}\label{sec:user-guide}

We will illustrate the main idea of the design methods though a simple motion planning example where the robot position in the workspace is uncertain. In this scenario, we assume that the robot localization depends on the amount of light at its current position as shown in \cref{fig:light-dark-example-workspace}. We can model the robot dynamics in the workspace plane, i.e., $\boldsymbol{x} \in \mathbb{R}^2$, as a first-order system controlled by velocity, $\boldsymbol{u} \in \mathbb{R}^2$: $\boldsymbol{x}_{k+1} = \boldsymbol{x}_k + 0.25 \boldsymbol{u}_k$. The observation function is the identity $\boldsymbol{y}_k = \boldsymbol{x}_k + n(\boldsymbol{x}_k) \boldsymbol{v}_k$ with a zero-mean Gaussian noise function as follows: 
     \begin{equation}
         n(\boldsymbol{x}) = 0.1(5 - x_1)^2 + const,
     \end{equation}
     where $\boldsymbol{x} = [x_1, x_2]^\intercal$. 

We do not know the robot's position in the workspace. However, our initial belief is an isotropic Gaussian distribution centered at position $[0, 2.5]^\intercal$ with covariance $diag(0.1, 0.1)$. 

We can specify the motion planning requirements as an \gls*{prstl} formula as follows:
 \begin{equation}
 \mFormula = safe \boldsymbol{U}_{[0,240]} \mAlways_{[0,40]} target,
 \end{equation}
 where $safe = \pi_{0.01}^{-x_1 - 1} \mAnd \pi_{0.01}^{x_1 - 5} \mAnd \pi_{0.01}^{-x_2 - 1} \mAnd \pi_{0.01}^{x_2 - 4}$ and $target = \pi_{0.05}^{-x_1 - 0.25} \mAnd \pi_{0.05}^{x_1 - 0.25} \mAnd \pi_{0.05}^{-x_2 - 0.25} \mAnd \pi_{0.05}^{x_2 - 0.25}$.
 In plain English, the robot must satisfy each safety boundary with $99\%$ of confidence until it achieves the target region with $95\%$ confidence within $240$ time instants and stays in the target for $40$ time instants.
 
 We can solve this problem using idSTLPy toolbox as shown in \cref{lst:light-dark}. In the subsequent sections, we will explain this code in more detail.

\begin{lstlisting}[language=Python,float=*,caption=Motion Planning Example,label=lst:light-dark, basicstyle=\ttfamily\scriptsize]
import idstlpy as stl
import numpy as np

q = stl.mk_variable(size=1, dtype=int)
x = stl.mk_variable(size=2, dtype=float)
u = stl.mk_variable(size=2, dtype=float)
problem = stl.Problem(
    switched_system=stl.mk_switched_sys([
        stl.dynamical_system.mk_lbs(
            A=np.identity(x.size), B=0.25 * np.eye(x.size, u.size), W=np.zeros((2, 2)),
            C=np.identity(2), V=lambda state: ((1 / 10) * (5 - state[0]) ** 2 + 0.001) * np.identity(2))
    ]),
    control_domain=stl.logical_and(u[0] >= -1.0, u[0] <= 1.0, u[1] >= -1.0, u[1] <= 1.0).region,
    initial_state=stl.to_belief(mean=np.array([0, 2.5]), cov=np.diag([0.1, 0.1])),
    stl_formula=stl.until(
        stl.logical_and(q == 0, stl.prob(x[0] >= -1) >= 1 - 0.01, stl.prob(x[0] <= 5) >= 1 - 0.01,
                                stl.prob(x[1] >= -1) >= 1 - 0.01, stl.prob(x[1] <= 4) >= 1 - 0.01, name='free_space'),
        0, 240,
        stl.always(0, 40, 
            stl.logical_and(q == 0, stl.prob(x[0] >= -0.25) >= 1 - 0.05, stl.prob(x[0] <= 0.25) >= 1 - 0.05,
                                    stl.prob(x[1] >= -0.25) >= 1 - 0.05, stl.prob(x[1] <= 0.25) >= 1 - 0.05,
                                    name='target')))
)
solution = problem.solve(rrt_timeout=60, delta_near=2, delta_drain=0.5, goal_bias=0.25, 
                         min_num_of_steps=3, max_num_of_steps=15)

for i in range(problem.switched_system.system_modes.size):
    problem.switched_system.system_modes[i].compute_finite_horizon_lqr(horizon=5, Q_final=np.identity(2), Q=np.identity(2), R=0.05*np.identity(2))
xi_sim, xi_real = problem.switched_system.simulate(
    reference_trajectory=solution, num_of_steps=solution.num_of_steps, real_initial_state=np.array([0.5, 2.75]),
    real_system=stl.mk_switched_sys([stl.mk_lcs(A=np.identity(2), B=0.25 * np.identity(2))])
)
\end{lstlisting}

\subsection{Systems}

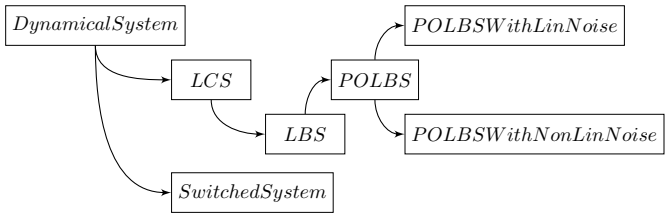
\begin{figure}
    \tikzstyle{block} = [draw, color=black, text=black, line width=0.1mm, rectangle, minimum height=2em, minimum width=4em]
    \centering
    \begin{tikzpicture}[auto, >=latex', scale=0.75, transform shape]
        \node[block] (dyn) {$DynamicalSystem$};
        \node[block, below right=0.25cm and -0.25cm of dyn] (lcs) {$LCS$};
        \node[block, below right=0.25cm and 0.25cm of lcs] (lbs) {$LBS$};
        \node[block, above right=0.25cm and -0.25 of lbs] (polbs) {$POLBS$};
        \node[block, above right=0.25cm and -0.25cm of polbs] (POLBSWithLinNoise) {$POLBSWithLinNoise$};
        \node[block, below right=0.25cm and -0.25cm of polbs] (POLBSWithNonLinNoise) {$POLBSWithNonLinNoise$};
        \node[block, below right=2.25cm and -0.25cm of dyn] (ss) {$SwitchedSystem$};
        \draw[->] (dyn.south) to[out=-90,in=180] (lcs.west);
        \draw[->] (dyn.south) to[out=-90,in=180] (ss.west);
        \draw[->] (lcs.south) to[out=-90,in=180] (lbs.west);
        \draw[->] (lbs.north) to[out=90,in=180] (polbs.west);
        \draw[->] (polbs.north) to[out=90,in=180] (POLBSWithLinNoise.west);
        \draw[->] (polbs.south) to[out=-90,in=180] (POLBSWithNonLinNoise.west);
    \end{tikzpicture}
    \caption{Class Diagram representation of dynamical system types and their inheritance. }
    \label{fig:dynamical-system-diagram}
    \vspace{-0.5cm}
 \end{figure}

A system (a.k.a., switched system) is composed of a finite set of system modes. Each system mode is also a dynamical system that inherit the behavior of a linear control system (LCS, i.e., $\boldsymbol{x}_{k+1} = A \boldsymbol{x}_k + B \boldsymbol{u}_k$) as illustrated in \cref{fig:dynamical-system-diagram}. We can have three types of dynamical behavior. If the output dimension is zero, i.e., $p = 0$, the system mode is a linear belief system (LBS). In this behavior, there is no active perception because we assume no observation. Otherwise, if the output dimension is non-zero (i.e., $p > 0$), we have a partially observable linear belief system (POLBS). In turn, a POLBS can have a linear noise function (POLBSWithLinNoise, i.e., $n(\boldsymbol{x}) = V$, where $V \in \mathbb{R}^{p \times p}$ is a constant matrix) or a nonlinear noise function (POLBSWithNonLinNoise). 

We can create any one of the linear belief systems using the function $mk\_lbs(A, B, W, C=None, V=None)$, where $C$ and $V$ are optional parameters, and $V$ can be a constant matrix or a function that returns a matrix. Similarly, we can construct a switched system using the function $sys.mk\_switched\_sys(system\_modes)$, where $system\_modes$ is a list of linear belief systems. We declare a dynamical system in our example in \cref{lst:light-dark} lines 8-12.

\subsection{Variables}

Due the hybrid nature of switched systems, we have two variable types: real-valued and discrete. As illustrated in \cref{fig:variable-diagram}, a real-valued variable is also a linear expression. A linear expression over a variable $\boldsymbol{x} \in \mathbb{R}^n$ is a multiplication between a constant vector $\boldsymbol{h} \in \mathbb{R}^n$ and the variable plus a constant $c \in \mathbb{R}$: $\boldsymbol{h}^\intercal \boldsymbol{x} + c$. For example, variable $\boldsymbol{x} \in \mathbb{R}^2$ is also a vector of linear expressions: $\big((1, 0)^\intercal \boldsymbol{x} + 0.0, (0, 1)^\intercal \boldsymbol{x} + 0.0\big)$. A discrete variable is a variable which can assume a finite set of values. For example, in \cref{lst:light-dark}, $q$ is a discrete variable that have one valid value (i.e., $q \in \{ 0 \}$). 

We can declare a variable by the function $mk\_variable(size, dtype)$, where $dtype$ is one of of these two types: $float$ for real-valued variables, $int$ for discrete variables. If $dtype$ is $float$, $size$ is the variable dimension. If $dtype$ is $int$, $size$ is the cardinality of the variable set of values. We declare variables in our example in \cref{lst:light-dark} lines 4-6.

  \begin{figure}
    \tikzstyle{block} = [draw, color=black, text=black, line width=0.1mm, rectangle, minimum height=2em, minimum width=4em]
    \centering
    \begin{tikzpicture}[auto, >=latex', scale=0.75, transform shape]
        \node[block] (expr) {$LinearExpression$};
        \node[block, right=1cm of expr] (rvar) {$RealValuedVariable$};
        \node[block, below=0.25cm of expr] (dvar) {$DiscreteVariable$};
        \draw[->] (expr) -- (rvar);
    \end{tikzpicture}
    \caption{Class Diagram representation of variable types and their inheritance. }
    \label{fig:variable-diagram}
    \vspace{-0.5cm}
 \end{figure}   
 
 \subsection{Constraints}
 
 An \gls*{prstl} formula atom is a constraint over the discrete and real-valued variables. We illustrate the inheritance of different constraint objects in \cref{fig:convex-constraints-diagram}. We call a constraint defined as an equality over discrete variables $q = \alpha$ as discrete predicate (DiscretePredicate implements $\pi^{\{\alpha\}}$), where $\alpha \in \mathbb{N}$. On the other hand, if the constraint is defined over a real-valued variable, this constraint is convex (ConvexConstraint). Particularly, a linear inequality over the variable (i.e., $\boldsymbol{h}^\intercal \boldsymbol{x} + c \leq 0$) is a linear predicate (LinearPredicate). Similarly, a inequality over the probability of a linear predicate (i.e., $prob(\boldsymbol{h}^\intercal \boldsymbol{x} + c \leq 0) >= 1 - \epsilon$) is a probabilistic linear predicate (ProbabilisticLinearPredicate implements $\pi^{\boldsymbol{h}^\intercal \boldsymbol{x} +c}_\epsilon$).
 
   \begin{figure}
    \tikzstyle{block} = [draw, color=black, text=black, line width=0.1mm, rectangle, minimum height=2em, minimum width=4em]
    \centering
    \begin{tikzpicture}[auto, >=latex', scale=0.75, transform shape]
        \node[block] (cc) {$ConvexConstraint$};
        \node[block, above right=0.25 and 0.25cm of cc] (lpi) {$LinearPredicate$};
        \node[block, right=0.5cm of lpi] (plpi) {$ProbabilisticLinearPredicate$};
        \node[block, below=1cm of cc] (dpi) {$DiscretePredicate$};
        \node[block, below right=0.25 and 0.25cm of cc] (region) {$ConvexRegion$};
        \draw[->] (lpi) -- (plpi);
        \draw[->] (dyn.north) to[out=90,in=180] (lpi.west);
        \draw[->] (dyn.south) to[out=-90,in=180] (region.west);
    \end{tikzpicture}
    \caption{Class Diagram representation of variable types and their inheritance. }
    \label{fig:convex-constraints-diagram}
    \vspace{-0.5cm}
 \end{figure}   
 
\begin{figure}
    \tikzstyle{block} = [draw, color=black, text=black, line width=0.1mm, rectangle, minimum height=2em, minimum width=4em]
    \centering
    \begin{tikzpicture}[auto, >=latex', scale=0.75, transform shape]
        \node[block] (poly) {$Polytope$};
        \node[block, above right=0.25cm and -0.25cm of dyn] (fullpoly) {$FullDimensionalPolytope$};
        \node[block, below right=0.25cm and 0.25 of poly] (belief_cone) {$BeliefCone$};
        \node[block, below right=0.25cm and -0.25cm of fullpoly] (full_belief_cone) {$FullDimensionalBeliefCone$};
        \draw[->] (poly.north) to[out=90,in=180] (fullpoly.west);
        \draw[->] (poly.south) to[out=-90,in=180] (belief_cone.west);
        \draw[->] (fullpoly.south) to[out=-90,in=180] (full_belief_cone.west);
    \end{tikzpicture}
    \caption{Class Diagram representation of polytope types and their inheritance. }
    \label{fig:polytope-diagram}
    \vspace{-0.5cm}
 \end{figure}
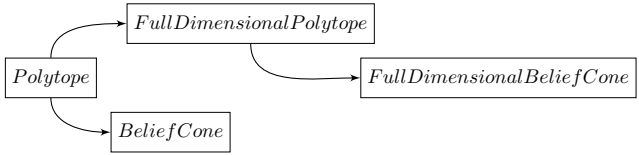

 We can apply the Boolean conjunction operator $\mAnd$ over linear and probabilistic linear predicates (i.e., the function $logical\_and(*args)$\footnote{In Python language, $*args$ means a variable number of arguments.}). The resulting constraint is a convex region over the state (if arguments are LinearPredicates) or belief state space (if arguments are ProbabilisticLinearPredicate). A convex region has a property that defines a polytope or a belief cone as illustrated in \cref{fig:polytope-diagram}. We declare a convex region and take its region that defines the input domain (a polytope) in our example in \cref{lst:light-dark} line 13. 
 
\subsubsection{Polytopes}

A \emph{polytope} $\mathcal{X} \subseteq \mathbb{R}^n$ is a set in $\mathbb{R}^n$ defined by the intersection of a finite number of closed half-spaces, i.e, $\mathcal{X} := \cap_i \{ \mu_i(\boldsymbol{x}) \leq 0 \}$, where $\mu_i \in H(\mathcal{P})$ is a linear function $\mu_i : \mathbb{R}^n \rightarrow \mathbb{R}$ such that $\mu_i(\boldsymbol{x}) := \boldsymbol{h}_i^\intercal \boldsymbol{x} + c_i$, $H(\mathcal{P})$ is the set of linear functions that defines the polytope $\mathcal{P}$, $\boldsymbol{h}_i \in \mathbb{R}^n$ and $c_i \in \mathbb{R}$ are constants. We can also represent a compact polytope $\mathcal{X} \subset \mathbb{R}^n$ as the convex hull of its vertices, i.e., $\mathcal{X} = \text{conv}\big(V(\mathcal{X})\big)$, where $V(\mathcal{X})$ is a set of vertices.

\subsubsection{Belief Cone}
We call the reciprocal of polytopes in the belief state space as belief cones. These cones $\mathcal{B} \subseteq \mathbb{R}^{n (n + 1)}$ are intersection of a finite set of second order cones about the multivariate Gaussian distribution parameters (i.e., mean $\hat{\boldsymbol{x}} \in \mathbb{R}^n$ and covariance $\Sigma^x \in \mathbb{R}^{n \times n}$) that satisfies a conjunction of a finite set of probabilistic linear predicates (i.e, $\bigwedge_i prob(\mu_i(\boldsymbol{x}) \leq 0) \geq 1 - \epsilon_i$). For simplicity, we will denote that a Gaussian random variable $\boldsymbol{X}_k \sim \mathcal{N}(\hat{x}_k, \Sigma_k^x)$ satisfy a probabilistic linear predicate $prob(\mu_i(\boldsymbol{x}) \leq 0) \geq 1 - \epsilon_i$ by $\boldsymbol{X}_k \mSat prob(\mu_i(\boldsymbol{x}) \leq 0) \geq 1 - \epsilon_i$. Therefore, a belief cone is defined as:
\begin{equation}
\begin{aligned}
     \mathcal{B} := & \{ \boldsymbol{b} \in \mathbb{R}^{n (n+1)}: \bigwedge_i \boldsymbol{X} \mSat prob(\boldsymbol{h}_i^\intercal \hat{\boldsymbol{x}}_k + c_i \leq 0) \geq 1 - \epsilon_i\} \\
     :=& \cap_i \{ \boldsymbol{h}_i^\intercal \hat{\boldsymbol{x}}_k + c_i + \Phi^{-1}(1 - \epsilon_i) \sqrt{\boldsymbol{h}_i^\intercal \Sigma_k^x \boldsymbol{h}_i}  \leq 0 \},
\end{aligned}
\end{equation}
where $\boldsymbol{b} \in \mathbb{R}^{n (n+1)}$ is the Gaussian distribution parameter variable, $\boldsymbol{h}_i^\intercal \in \mathbb{R}^n$, $c_i \in \mathbb{R}$ and $\epsilon_i \in [0, 0.5]$ are constants, $\Phi(v)$ and  $\Phi^{-1}(p)$ are the cumulative distribution and quantile functions of the standard Gaussian distribution $V \sim \mathcal{N}(0,1)$, i.e., $\Phi(v) = prob(V \leq v)$ and $\Phi^{-1}(p) \leq v$ if and only if $p \leq \Phi(v)$. We can easily see that $\boldsymbol{X} \mSat prob(\boldsymbol{h}_i^\intercal \hat{\boldsymbol{x}}_k + c_i \leq 0) \geq 1 - \epsilon_i$ if and only if $\boldsymbol{h}_i^\intercal \hat{\boldsymbol{x}}_k + c_i + \Phi^{-1}(1 - \epsilon_i) \sqrt{\boldsymbol{h}_i^\intercal \Sigma_k^x \boldsymbol{h}_i}  \leq 0$ from Gaussian distribution properties such as linear transformation and the quantile function definition.

\subsection{PrSTL formula}

\begin{figure}
    \tikzstyle{block} = [draw, color=black, text=black, line width=0.1mm, rectangle, minimum height=2em, minimum width=4em]
    \centering
    \begin{tikzpicture}[auto, >=latex', scale=0.75, transform shape]
        \node[block] (phi) {$STLFormula$};
        \node[block, above right=0.25cm and 0.25cm of phi] (tphi) {$STLTemporalFormula$};
        \node[block, below right=0.25cm and 0.25 of phi] (pi) {$STLAtomicProposition$};
        \node[block, above left=0.25cm and 0.25cm of phi] (phi_and) {$STLAnd$};
        \node[block, below left=0.25cm and 0.25cm of phi] (phi_or) {$STLOr$};
        \node[block, above right=0.25cm and 0.25cm of tphi] (phi_until) {$STLUntil$};
        \node[block, below right=0.25cm and 0.25cm of tphi] (phi_release) {$STLRelease$};
        \draw[->] (phi.north) to[out=90,in=180] (tphi.west);
        \draw[->] (tphi.north) to[out=90,in=180] (phi_until.west);
        \draw[->] (phi.south) to[out=-90,in=180] (pi.west);
        \draw[->] (phi.north) to[out=90,in=0] (phi_and.east);
        \draw[->] (phi.south) to[out=-90,in=0] (phi_or.east);
        \draw[->] (tphi.south) to[out=-90,in=180] (phi_release.west);
    \end{tikzpicture}
    \caption{Class Diagram representation of \gls*{prstl} formula types and their inheritance. }
    \label{fig:prstl-diagram}
    \vspace{-0.5cm}
 \end{figure}
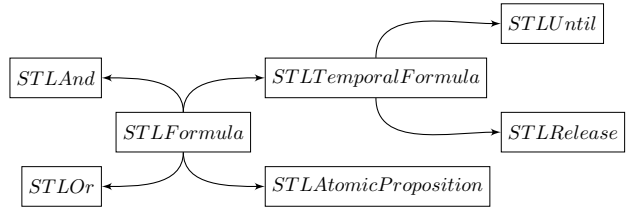

We implement a \gls*{prstl} formula as classes shown in \cref{fig:prstl-diagram}. An STLAtomicProposition implements a \gls*{prstl} state formula $\phi$, meaning that it represents the conjunction (i.e., $logical\_and(*args)$) over a list of ProbabilisticLinearPredicate and Boolean formula (i.e., using $logical\_and(*args)$ or $logical\_or(*args)$) over DiscretePredicate, meaning that it is defined by a convex region and a set of valid system modes. In plain English, a trajectory $\boldsymbol{\xi}$ satisfies an STLAtomicProposition at instant $k$ if it reaches a belief state in the convex region using one of the valid system modes. We declare two STLAtomicProposition in our example in \cref{lst:light-dark} lines 16-17 and lines 20-21. 

An STLAnd object represent the conjunction (i.e., $logical\_and(*args)$) of a list of STLFormulas containing at most one STLAtomicProposition. On the other hand, an STLOr object is a disjunction (i.e., $logical\_or(*args)$) of a list STLFormulas but containing an arbitrary number of STLAtomicProposition. An STLUntil object is an \gls*{prstl} formula with until operator $\mFormula_1 \mUntil_{[a,b]} \mFormula_2$, and STLRelease object is an \gls*{prstl} formula with until operator $\mFormula_1 \mRelease_{[a,b]} \mFormula_2$. We obtain these formulas using the functions: $until(\mFormula_1, a, b, \mFormula_2)$, $eventually(a, b, \mFormula)$, $release(\mFormula_1, a, b, \mFormula_2)$, $always(a, b, \mFormula)$. We declare a formula in our example in \cref{lst:light-dark} lines 15-22. 

\subsection{Solution}

The object Problem wraps the implementation of the approach presented in \cref{sec:overview}. A solution could be empty ($None$ in Python) if the algorithm did not find a trajectory that satisfies the specification. However, if such a trajectory is found, the algorithm returns an object that implements this trajectory. Since the solution is a trajectory of the approximated transition system \cite{dasilva2021active}, the returned trajectory is a ProbabilisticTSTrajectory. This object implements methods to extract data from the trajectory such as: $get\_mean$, $get\_cov$, $get\_control$, $get\_action$ that returns the belief state mean, the belief state covariance, the control, and the action from the trajectory. We can use these methods to execute a trajectory tracking strategy. We declare a problem in our example in \cref{lst:light-dark} lines 7-23 and solve it in \cref{lst:light-dark} lines 24-25.

We can simulate the execution of this planned trajectory. The planned trajectory is the result of an approximated belief dynamics where the observations are the \gls*{mlo}. Hence, we propose to use a linear feedback law to adjust the belief state during the execution to track the planned trajectory. Specifically, we implemented a \gls*{rhc} strategy with a finite horizon discrete time \gls*{lqr} to track the belief mean values. We initialize this strategy by calling the method $compute\_finite\_horizon\_lqr(horizon, Q\_final, Q, R)$ for each system mode, where $Q_{final}, Q \in \mathbb{R}^{n \times n}$ are positive-semi-definite contrant matrices and $R \in \mathbb{R}^{m \times m}$ is a positive definite matrix. Next, we call the method $simulate$ from SwitchedSystem object. This method simulates the system $real\_system$ initialized at $real\_initial\_state$ for $num\_of\_steps$ steps while using the linear feedback law to track the planned trajectory $reference\_trajectory$. In the running example, we track the mean of the estimated belief with cost function $J = \hat{\boldsymbol{x}}_h^\intercal Q_{final} \hat{\boldsymbol{x}}_h + \sum_{k=0}^{h - 1} \hat{\boldsymbol{x}}_k^\intercal Q \hat{\boldsymbol{x}}_k + \boldsymbol{u}_k^\intercal R \boldsymbol{u}_k$, where $Q = I_2$ and $R = 0.05 I_2$ and the horizon $h = 5$, as shown in \cref{lst:light-dark} lines 27-32. 

\begin{figure}
        \centering
        \includegraphics[width=0.9\linewidth]{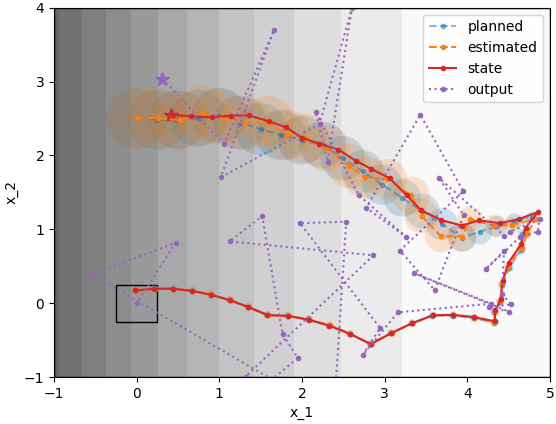}
        \caption{Light-dark example. The shade in the workspace represents the amount of light at that position and the black box is the target.}
        \label{fig:light-dark-example-workspace}
        \vspace{-0.5cm}
    \end{figure}

The result is shown in \cref{fig:light-dark-example-workspace}. The blue trajectory is the planned trajectory in the belief space. This trajectory approximates the observation as maximum likelihoods. However, we observe a very different observation during the execution, which is the purple trajectory in the figure. As a result, the belief trajectory during the execution is slightly different, the orange trajectory in the figure. However, this trajectory satisfies the specification, and the resulting state (in red) also is within the expected result. Since the maximum likelihood approximation is close to the most likely belief trajectory, a simple tracking strategy is, in general, sufficient to enforce the planned trajectory during execution.

\section{Conclusions and future work}\label{sec:conclusion}

In this work, we presented a Python toolbox for controller synthesis from PrSTL specifications. We considered problems with a switched linear system with Gaussian noises. We illustrated our approach on a simulation of robot motion planning under noisy localization. In this example, we showed that the planned trajectory satisfied both the task and active perception requirements.


We will focus on two directions in future work. One direction is to extend to other probabilistic hybrid systems and also consider probabilistic switching. Another direction is to drop the \gls*{mlo} approximation during the planning without a conservative assumption. 

\bibliographystyle{IEEEtran}
\bibliography{IEEEabrv,library} 

\vspace{12pt}

\end{document}